# Optimization of heterogeneous ternary $Li_3PO_4$-$Li_3BO_3$-$Li_2SO_4$ mixture for Li-ion conductivity by machine learning

Kenji Homma[*, §], Yu Liu[§], Masato Sumita[†, #], Ryo Tamura[†, #, ‡, ¶],
Naoki Fushimi[§], Junichi Iwata[§], Koji Tsuda[†, ‡, ¶], Chioko Kaneta[§]

*Corresponding author. E-mail: K.H., homma.kenji@fujitsu.com
[§]*Fujitsu Laboratories Ltd. 10-1 Morinosato-Wakamiya, Atsugi 243-0197, Japan.*
[†] *Centre for Advanced Intelligence Project, RIKEN, 1-4-1 Nihombashi, Chuo-ku, Tokyo, 103-0027, Japan.*
[#]*International Center for Materials Nanoarchitectonics (WPI-MANA), National Institute for Materials Science, 1-1 Namiki, Tsukuba, Ibaraki, 305-0044, Japan.*
[‡]*Graduate School of Frontier Sciences, The University of Tokyo, 5-1-5 Kashiwa-no-ha, Kashiwa, Chiba, 277-8561, Japan.*
[¶]*Research and Services Division of Materials Data and Integrated System, National Institute for Materials Science, 1-2-1 Sengen, Tsukuba, Ibaraki, 305-0047, Japan.*

**Abstract**
Mixing heterogeneous Li-ion conductive materials is one of potential ways to enhance the Li-ion conductivity more than that of the parent materials. However, the development of the mixtures had not exhibited significant progress because it is a formidable task to cover the vast possible composition of the parent materials using traditional ways. Here, we introduce a fashion based on machine learning to optimize the composition ratio of ternary $Li_3PO_4$-$Li_3BO_3$-$Li_2SO_4$ mixture for its Li-ion conductivity. According to our results, the optimum composition of the ternary mixture system is 25:14:61 ($Li_3PO_4$: $Li_3BO_3$: $Li_2SO_4$ in mol%), whose Li-ion conductivity is measured as $4.9 \times 10^{-4}$ S/cm at 300 °C. Our X-ray structure analysis indicates that Li-ion conductivity in the mixing systems is enhanced by virtue of the coexistence of two or more phases. Although the mechanism enhancing Li-ion conductivity is not simple, our results demonstrate the effectiveness of machine learning for the development of materials.

## 1. Introduction

Although Li-ion conductive oxides are attractive materials as an electrolyte of Li-ion secondary batteries because of their stability and low toxicity, their low Li-ion conductivity hinders their practical use.[1-3] Ordinal strategy to elevate the Li-ion conductivity of materials is to dope additives or mix heterogeneous materials.[4-10] These manners always have a common formidable problem, that is, to find the optimum composition. Covering all composition is impossible because of the vast search space. Recently, machine learning is often employed to address the problem.

The Bayesian optimization (BO), which is one of the machine learning techniques, makes it possible to find the optimum composition within the number of trials as small as possible.[11] In materials science, many successful examples with BO have been recently reported using both experiments and simulations for not only compositional optimization but also structural and process optimizations.[12,13] The BO can efficiently recommend a candidate composition of a mixed material process possessing the desired properties in spite of limited materials data.

In this study, using experimental measurements with BO, we perform optimization of the Li-ion conductivity in the ternary solid-oxide electrolyte system, $Li_3PO_4$-$Li_3BO_3$-$Li_2SO_4$. This mixed material is stable under atmospheric conditions and can be sintered integrally with the positive and negative electrode materials using ceramic inductor manufacturing technology. In the simulation based study, we have predicted that the theoretical $Li_3PO_4$-$Li_3BO_3$-$Li_2SO_4$ ternary system would be more Li-ion conductive than the parent materials because the parent materials possess poly-anions whose ionic valence are different each other.[14] However, in our theoretical prediction, it is difficult to identify the optimum composition with high Li-ion conductivity because of several computational limitations. This is the first report from the aspect of experiment for the $Li_3PO_4$-$Li_3BO_3$-$Li_2SO_4$ ternary systems though its binary parts have already been reported[15-17]. We have succeeded to obtain the optimum polycrystalline material with composition of 25:14:61 (mol%) for $Li_3PO_4$-$Li_3BO_3$-$Li_2SO_4$ which exhibits three times Li-ion conductivity ($4.9 \times 10^{-4}$ S/cm) compared with the highest case in binary parts.

Conventionally, material scientists try to analyze the crystal structure because the information is supposed to be beneficial for material design or improvement. Thus, we also try to elucidate the crystal structure of the mixed system. The result of X-ray diffraction pattern suggests that each crystal phase of parent materials ($Li_3PO_4$, $Li_3BO_3$, and, $Li_2SO_4$) and unknown phases coexist in the optimum polycrystalline material. Although it is difficult to clarify the effective conduction path of Li ions by only X-ray diffraction pattern, mixture of phases and grain boundaries among them should be related to it. Our results show that the synthesis method coupled with machine learning is effective for the optimization of material characteristics, which is an alternative method of conventional synthesis ones based on trial and error using human knowledge.

## 2. Experimental and machine learning methods

Polycrystalline samples were prepared using solid-state reaction methodologies. $Li_3PO_4$ (Kojundo Chemical Laboratory, >99.9% purity), $Li_3BO_3$ (Kojundo Chemical Laboratory, >99.99% purity), and $Li_2SO_4$ (Aldrich, >99% purity) were used as the parent materials. Samples were made by mixing the parent materials weighed with the appropriate molar ratios and sintered at appropriate temperature, which was determined from the measurements of melting points and phase transitions of the mixtures of the parent materials (620 – 1000 °C range). The melting points were measured with differential thermal analyzer (Rigaku, Thermo plus EVO2 TG 8120) at a heating rate of 10 °C/min in argon flow. After sintering, the samples were cooled down to room temperature over 12 hours in argon flow.

The ionic conductivity was measured using an AC impedance method under an argon atmosphere at 300 °C using a frequency response analyzer (Bio-Logic, VMP-300) with an applied frequency range of 1 MHz to 100 Hz. Gold thin films used for the blocking electrode were deposited on either side of the sample. X-ray diffraction patterns of samples were obtained using an X-ray diffractometer (Rigaku, Miniflex 600) with Cu$K\alpha$ radiation, and the data was collected at 0.02° intervals over a $2\theta$ range from 10° to 90°.

Using experimentally measured Li-ion conductivity for each ternary composition as the training dataset, the machine learning model is trained to predict the Li-ion conductivity. Here, to select the next candidate based on the BO context, the Gaussian process implemented in COMBO package[18, 19] is used. In COMBO package, the hyperparameters used in the Gaussian process are automatically searched. Furthermore, we employed slider space search method together with whole space searching to suggest next composition. In slider search method, the search space is limited to a line drawn by three known points with largest Li-ion conductivity in each binary system. In each iteration by BO, six compositions are simultaneously suggested by two acquisition functions (Maximum Expected Improvement (EI) and Maximum Probability of Improvement (PI)) and Thompson sampling in the both slider space and whole space, and the Li-ion conductivities of the suggested compositions are experimentally identified. The suggested materials are added into the training dataset, and the Gaussian process is retrained. By repeating this procedure, we could find the optimum composition with higher Li-ion conductivity even if the number of synthesized samples is small. Note that in our experimental design, the ternary composition was discretized in 1% increments.

## 3. Results and Discussion
### 3.1 Optimizing composition ratio for Li-ion conductivity with machine learning

We prepared 15 samples with the interval of 25% composition ratio in the analogous way to our simulation[14]. The Li-ion conductivities for the composition ratios of the initial samples are shown in Table 1. In order to visualize the correlation of the initial samples on the Li-ion ternary composition phase diagram, their Li-ion conductivity contour map interpolated among the samples are shown in Figure 1. This map indicates relatively high conductivity area lies at around the composition ratios of samples 7 (25:0:75 in mol%) and 8 (25:25:50 in mol%) whose Li-ion conductivities are respectively measured as $1.6 \times 10^{-4}$ S/cm and $1.7 \times 10^{-4}$ S/cm. This experimental tendency, which was obtained employing appropriate synthesis conditions for the individual samples, is different from the theoretical one[14] obtained by assuming limited conditions on crystal structure for each composition due to the available computational load. Therefore, the simple hypothesis we suggested in Ref. 14 might be only one of the factors to elevate the Li-ion conductivities the mixed system.

Starting with 15 known samples, we perform two iterations in BO. In our BO, 12 samples are recommended, though the same sample is suggested by different search methods in the second iteration. Thus, the 25 samples (Li-ion conductivities of different compositions) are totally synthesized. Here, the slider space in our problem is shown as the red line in Figure 1. The compositions of the additional samples suggested by BO are tabulated in Table 2, and samples 16 – 21 are recommended in first iteration of BO while the others are added in second iteration. In other words, in the first iteration, the optimum composition (e.g. sample 19) is found. Furthermore, the Li-ion conductivity distribution interpolated by Gaussian process using all measured data is shown in Figure 2. We find the limited region in which the ionic conductivity is improved. The maximum Li-ion conductivity is $4.9 \times 10^{-4}$ S/cm at sample 19 (25:14:61 in mol%), which is three times higher than the maximum conductivity observed in binary parts ($1.6 \times 10^{-4}$ S/cm).

**Table 1.** Initial samples whose composition ratios ($Li_3PO_4$, $Li_3BO_3$, $Li_2SO_4$) in percentage and their Li-ion conductivities measured at 300 °C.

| Sample No. | Ratio ($Li_3PO_4$, $Li_3BO_3$, $Li_2SO_4$) | Ionic conductivity S/cm |
|---|---|---|
| 1 | (100, 0, 0) | $3.9 \times 10^{-9}$ |
| 2 | (75, 0, 25) | $3.3 \times 10^{-5}$ |
| 3 | (75, 25, 0) | $1.7 \times 10^{-5}$ |
| 4 | (50, 0, 50) | $9.9 \times 10^{-5}$ |
| 5 | (50, 25, 25) | $9.7 \times 10^{-5}$ |
| 6 | (50, 50, 0) | $5.6 \times 10^{-7}$ |
| 7 | (25, 0, 75) | $1.6 \times 10^{-4}$ |
| 8 | (25, 25, 50) | $1.7 \times 10^{-4}$ |
| 9 | (25, 50, 25) | $8.3 \times 10^{-5}$ |
| 10 | (25, 75, 0) | $1.2 \times 10^{-6}$ |
| 11 | (0, 0, 100) | $1.4 \times 10^{-7}$ |
| 12 | (0, 25, 75) | $4.9 \times 10^{-5}$ |
| 13 | (0, 50, 50) | $2.3 \times 10^{-5}$ |
| 14 | (0, 75, 25) | $2.3 \times 10^{-5}$ |
| 15 | (0, 100, 0) | $9.1 \times 10^{-6}$ |

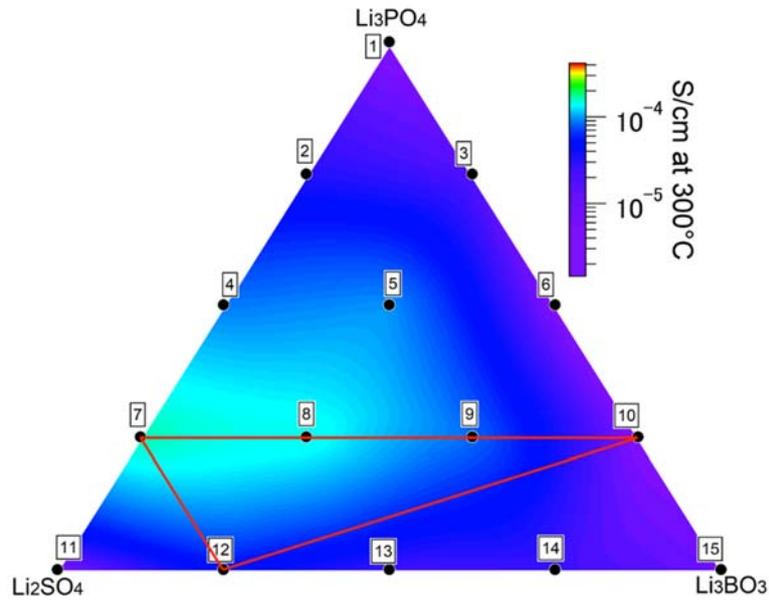

**Figure 1.** Ternary component contour map of Li-ion conductivity depicted by interpolating the data of the 15 initial samples shown in Table 1. Here, the Gaussian process by COMBO package is used to interpolate among Li-ion conductivities. Red line is the slider space where samples 7, 10, and 12 are compositions with largest Li-ion conductivity in each binary system.

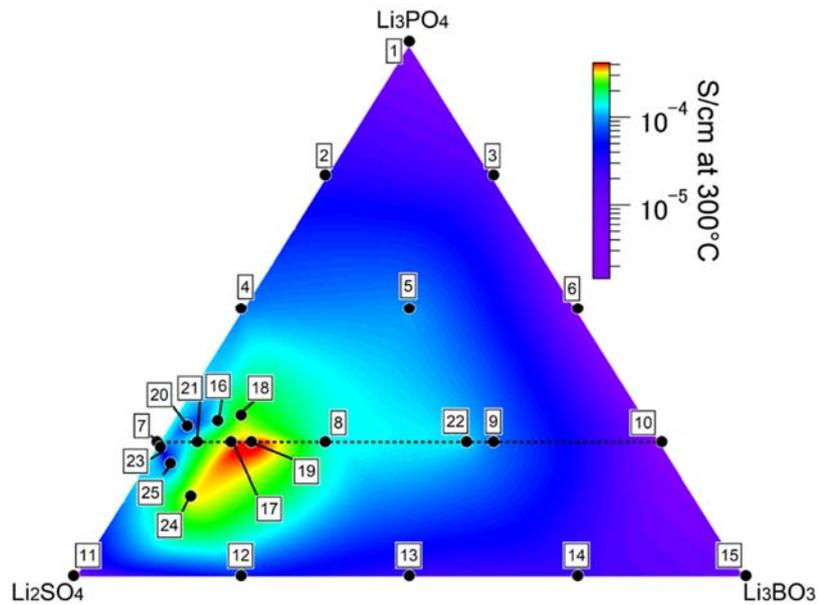

**Figure 2.** Ternary component contour map of Li-ion conductivity depicted by interpolating the data of all 25 samples shown in Tables 1 and 2. Here, the Gaussian process by COMBO package is used to interpolate of Li-ion conductivity. The broken line indicates the profile we focus on with X-ray analysis.

**Table 2.** Recommended samples by BO whose composition ratios (Li$_3$PO$_4$, Li$_3$BO$_3$, Li$_2$SO$_4$) in percentage and their ionic conductivity measured at 300°C.

| Sample No. | Ratio (Li$_3$PO$_4$, Li$_3$BO$_3$, Li$_2$SO$_4$) | Ionic conductivity S/cm |
|---|---|---|
| 16 | (29, 7, 64) | $1.3 \times 10^{-4}$ |
| 17 | (25, 11, 64) | $3.4 \times 10^{-4}$ |
| 18 | (30, 10, 60) | $2.6 \times 10^{-4}$ |
| 19 | (25, 14, 61) | $4.9 \times 10^{-4}$ |
| 20 | (28, 3, 69) | $4.9 \times 10^{-5}$ |
| 21 | (25, 6, 69) | $1.9 \times 10^{-4}$ |
| 22 | (25, 46, 29) | $1.2 \times 10^{-4}$ |
| 23 | (24, 1, 75) | $3.0 \times 10^{-5}$ |
| 24 | (15, 10, 75) | $3.3 \times 10^{-4}$ |
| 25 | (21, 4, 75) | $6.1 \times 10^{-5}$ |

### 3.2 X-ray structure analysis

Elucidating the crystal structure of Li-ion conductive materials had been supposed to be beneficial for improving Li-ion conductivities because the obvious crystal structure makes it possible to analyze Li-ion path. Hence, some theoretical and experimental researches had been performed to clarify the Li-ion path and evaluate the height of barrier for Li-ion migration.[20-23] However, for the heterogeneous mixed system like the target of this research, it is difficult to obtain the information only from the crystal structure. On the other hand, obtaining information of crystal structure of the mixed system might be beneficial to prove our hypothesis, in which the Li-ion conductivity of mixed system is enhanced by mixing heterogeneous materials whose poly-anions have different ionic valence each other.

Hereafter, we focus on the mixed systems on the broken line between samples 7 and 10 shown in Figure 2, which is able to be described in the formula, Li$_{2.25+2.33\delta}$B$_\delta$S$_{0.75-\delta}$P$_{0.25}$O$_{4-\delta}$. The X-ray diffraction peaks of the mixed systems and the parent materials: sample 1(γ-Li$_3$PO$_4$), sample 11(β-Li$_2$SO$_4$), sample 15 (α-Li$_3$BO$_3$) are shown in Figure 3. The profile of the Li-ion conductivities along the broken line in Figure 2 is summarized in Figure 4, which clearly indicates that sample 19 has maximum conductivity.

First, we focus on the difference between samples 7 and 21. Additional peaks in sample 21 from 7 suggest that a new phase is mixed into the parent materials: β-Li$_2$SO$_4$ and γ-Li$_3$PO$_4$, because the crystal phase of the sample 7 was a mixture of the two parent materials. From the exotic peak positions, we expect that these peaks are originated from a new Li$_2$SO$_4$ phase. Here, we call it γ-phase. Based on the X-ray diffraction pattern of the sample 21, the unit cell parameters of the γ-Li$_2$SO$_4$ were calculated by using the lattice calculation program, DICVOL06 (dichotomy method[24]). The dimensions of the monoclinic unit cell were obtained to be $a$ = 8.5509 Å, $b$ = 4.8321 Å, and $c$ = 15.9706 Å, $\beta$ = 99.4° (see the Supporting information). Here, the structure was analyzed in the space group $P2$ employing the direct space method of the FOX program[25]. In our Rietveld analysis, R$_{wp}$ was smaller than 10%, and the exotic peaks in the sample 21 would be came from the crystal phase of the γ-Li$_2$SO$_4$.

The X-ray analysis of sample 19 shows inherent peaks in Figure 3 in addition to the peaks observed in the sample 21, that is, at 2$\theta$ values of 7.6°, 23.3°, 26.0°, 27.0°, 33.2°, 35.7°, and 36.2° which are denoted by inverse triangles in Figure 3. Thus, a further new phase originated with these peaks is mixed in addition to β-Li$_2$SO$_4$, γ-Li$_3$PO$_4$, and γ-Li$_2$SO$_4$. However, we could not identify the crystal structure of this phase only from the X-ray diffraction pattern. On the other hand, in the materials with higher Li-ion conductivity, at least four or more phases including new ones and grain boundaries among them are mixed. Although we could not specify the origins of the higher Li-ion conductivity in above mentioned complicated systems, the formation of such mixed heterogeneous phases might be one of the factors improving the conductivity.

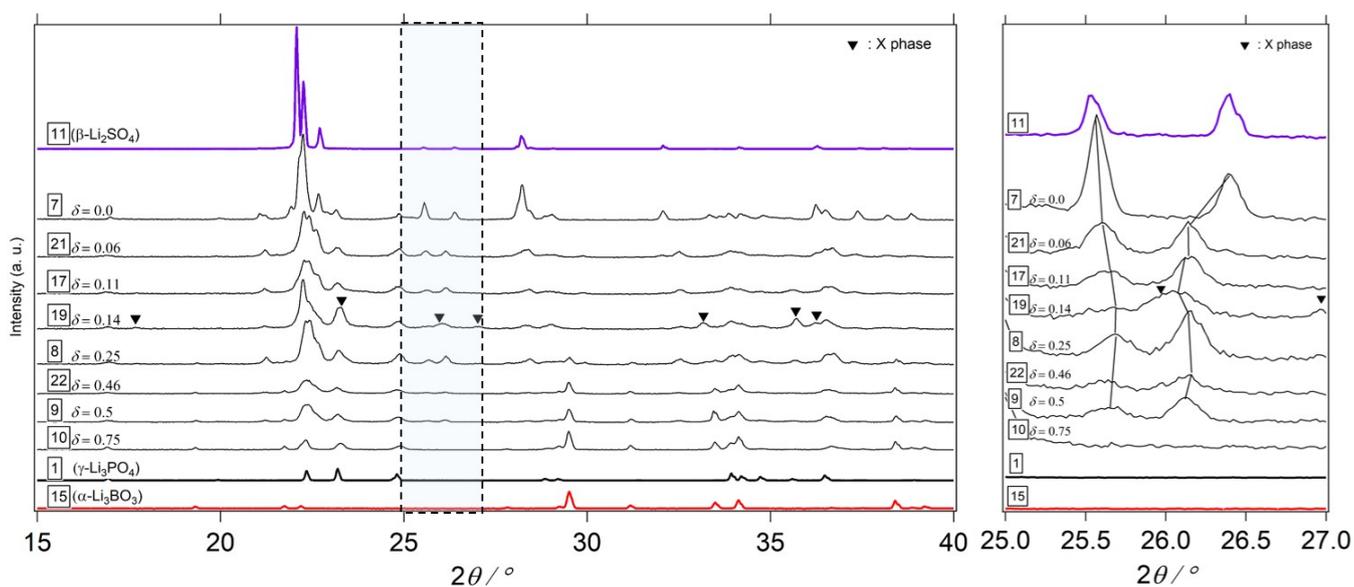

**Figure 3.** X-ray diffraction patterns on the line from samples 7 to 10 in Figure 2 and parent materials, sample 1($\gamma$-Li$_3$PO$_4$), sample 11($\beta$-Li$_2$SO$_4$), sample 15 ($\alpha$-Li$_3$BO$_3$). Right panel is enlarged view of the shaded area of left panel. Inverse triangles indicate inherent diffraction peaks of sample 19.

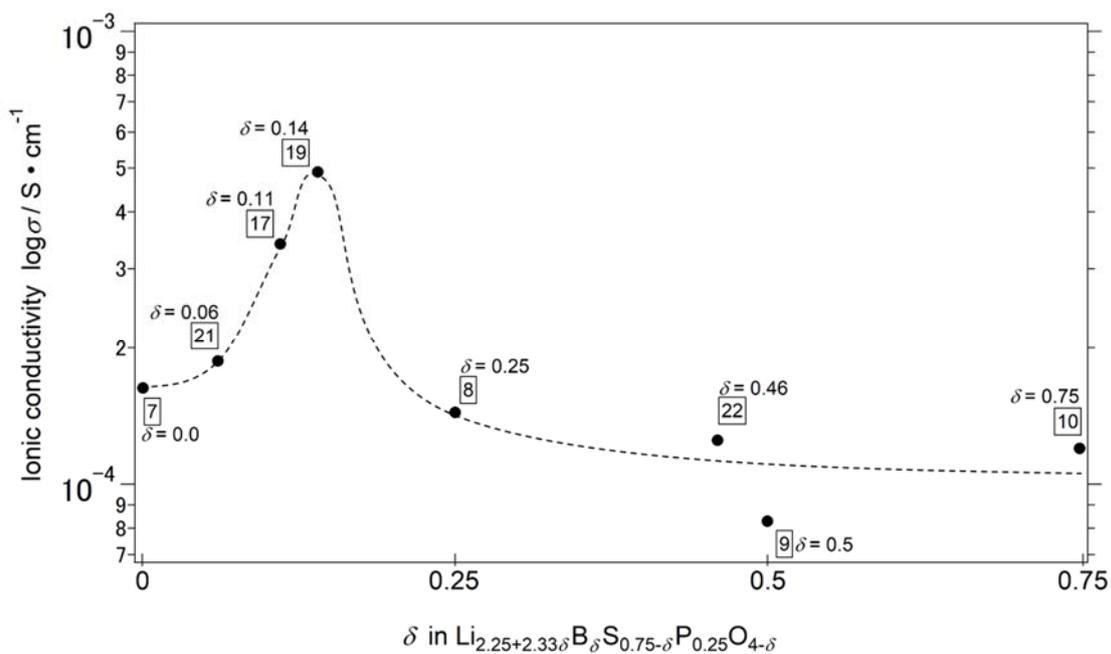

**Figure 4.** Ionic conductivity as the function of boron concentration ($\delta$) on the broken line from samples 7 to 10 (Li$_{2.25+2.33\delta}$B$_\delta$S$_{0.75-\delta}$P$_{0.25}$O$_{4-\delta}$) in Figure 2.

## 4. Conclusion

We have demonstrated the effectiveness of BO coupled with synthesizing and experimental measurement for compositional optimization of the Li-ion conductivity in the $Li_3PO_4$-$Li_3BO_3$-$Li_2SO_4$ ternary system. In the process of the BO without the bias of human, the composition of the $Li_3PO_4$-$Li_3BO_3$-$Li_2SO_4$ ternary system is optimized to 25 : 14 : 61 in mol% for the Li-ion conductivity which is $4.9 \times 10^{-4}$ S/cm. This value is three times higher than the maximum conductivity observed in binary parts ($1.6 \times 10^{-4}$ S/cm).

Our X-ray structure analysis have shown that some phases are mixed in the $Li_3PO_4$-$Li_3BO_3$-$Li_2SO_4$ ternary system. At least, we have found that two new phases in addition to parent materials are involved in optimum system. We have succeeded to confirm that one of two new phases is $\gamma$-$Li_2SO_4$, the other is still remain unknown. Although we could not locate the dominant factor improving Li-ion conductivity, these new phases and the grain boundaries among them might be related to the improvement of the Li-ion conductivities.

When the optimization of materials properties with experimental measurements is performed by BO, experimental time becomes large bottleneck in general. Thereby, iteration task in BO is often impossible in realistic time scale. However, our experiments using solid state reaction as synthesis method and AC impedance measurement can be performed in a relatively short time, and the optimization of Li-ion conductivity can be realized in the context of BO. Thus, we believe that the high Li-ion conductive materials will be found in the mixed materials with aid of machine learning.

**Supporting Information**

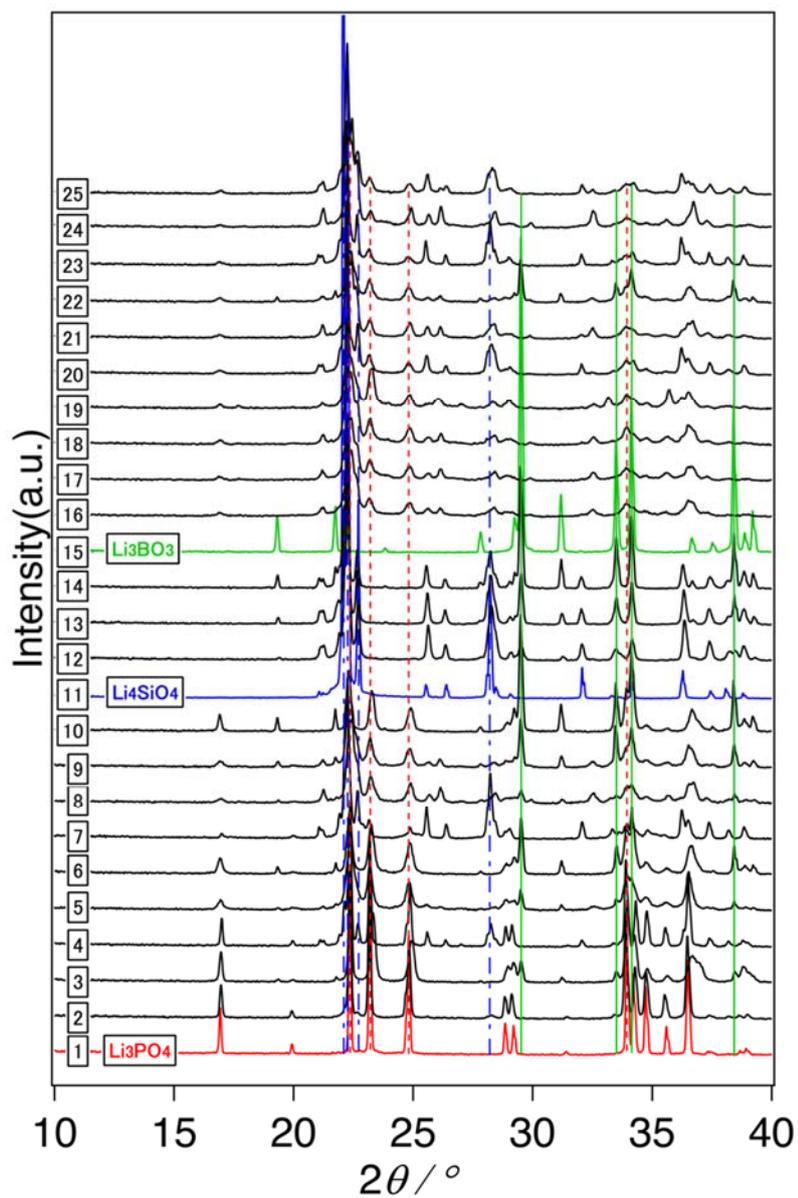

**Fig. S1** X-ray diffraction patterns of all samples (2θ = 15 - 40°). Samples 1, 11, and 15 correspond to the starting materials, γ-$Li_3PO_4$, β-$Li_2SO_4$, and α-$Li_3BO_3$, respectively. The diffraction patterns of the compounds have the peaks attributed to the parent materials. This fact indicates that two or more crystal phases coexist in the mixed systems.

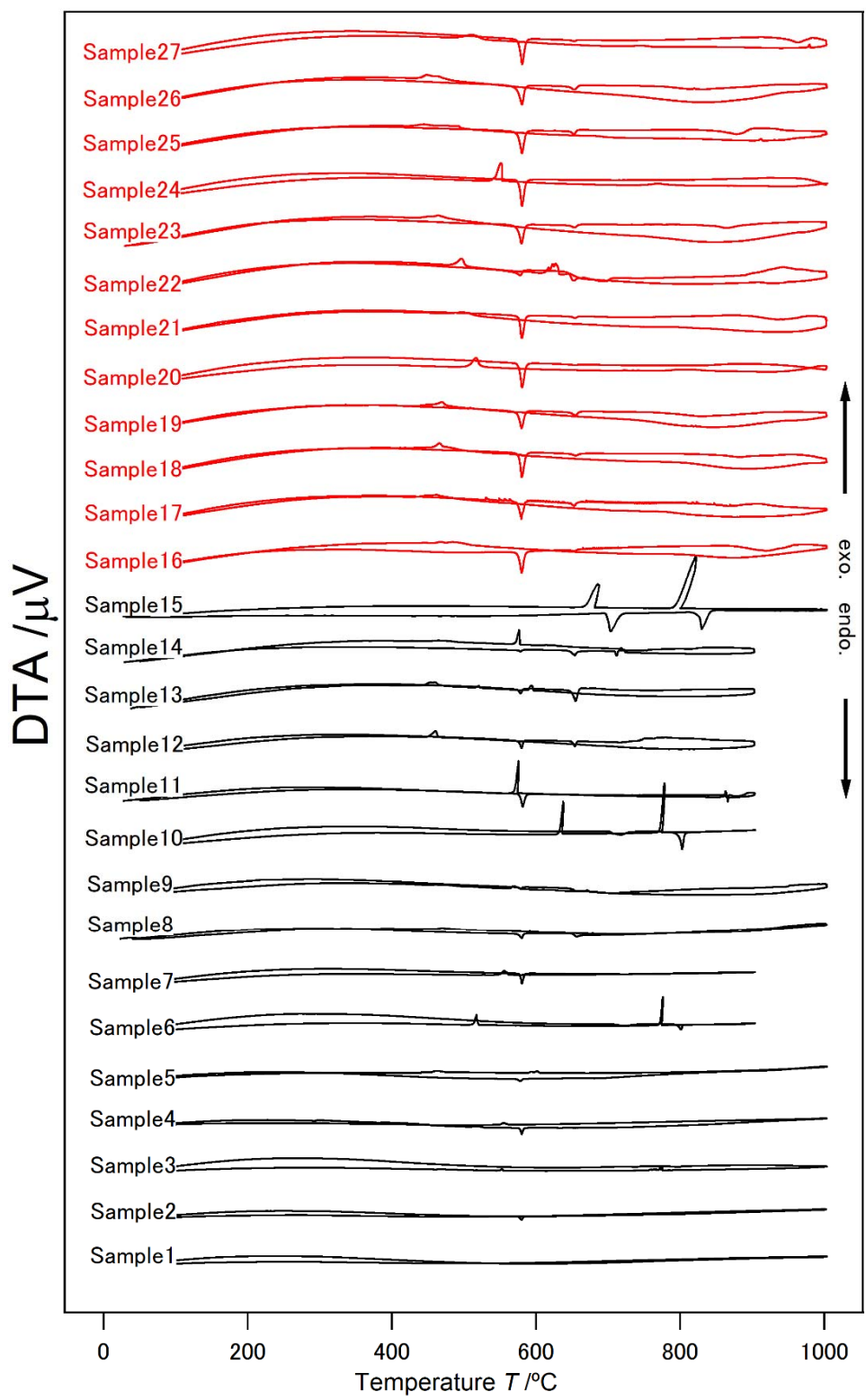

**Fig. S2** Differential thermal analysis curves of all samples.

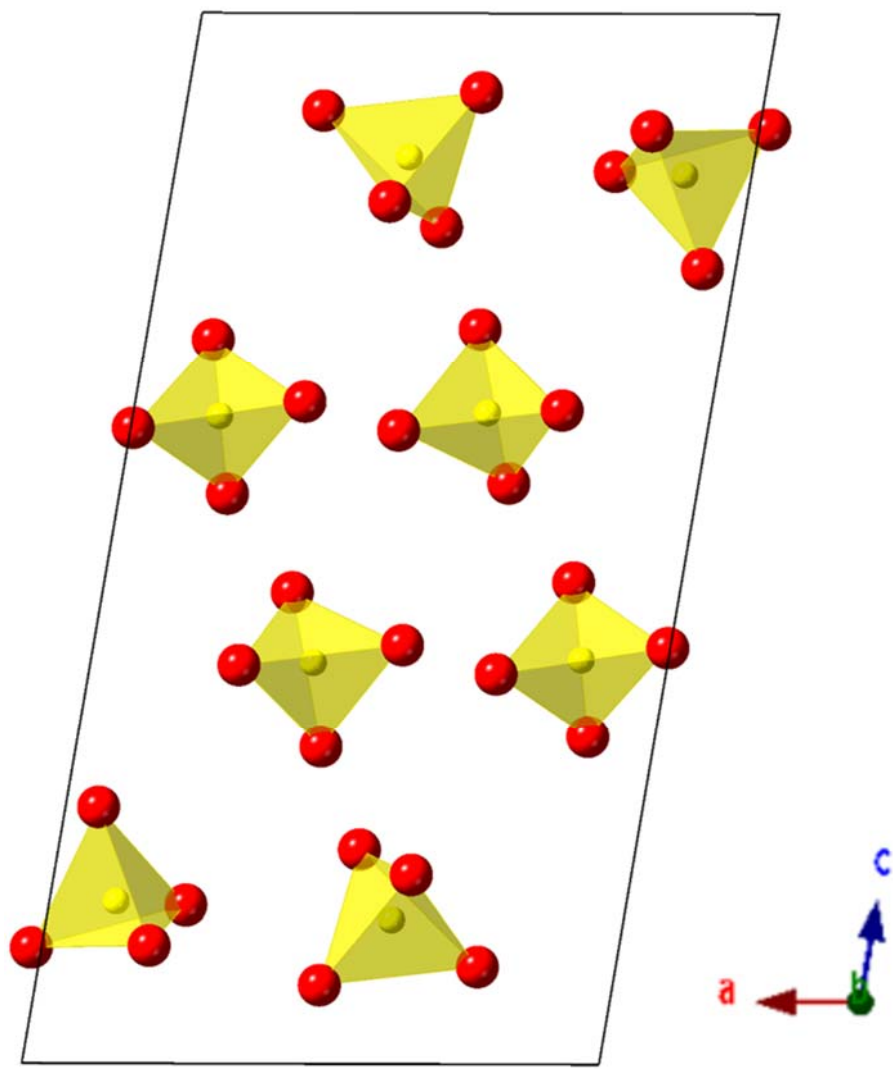

**Fig. S3** Crystal structure of γ-Li₂SO₄ projected along the *b* axis. The structural framework composed of a SO₄ tetrahedral unit.

**Table S1** Structural parameters of γ-Li$_2$SO$_4$ determined by direct space method.

| Atom | Site | g | x | y | z |
|---|---|---|---|---|---|
| O (1) | 2e | 1 | 0.7421 | 0.9840 | 0.3785 |
| O (2) | 2e | 1 | 0.6712 | 0.5678 | 0.4479 |
| O (3) | 2e | 1 | 0.4647 | 0.9165 | 0.3979 |
| O (4) | 2e | 1 | 0.5756 | 0.6270 | 0.3007 |
| O (5) | 2e | 1 | 0.9958 | 0.4734 | 0.6042 |
| O (6) | 2e | 1 | 0.8131 | 0.1236 | 0.5419 |
| O (7) | 2e | 1 | 0.8837 | 0.1172 | 0.6892 |
| O (8) | 2e | 1 | 0.7060 | 0.4823 | 0.6296 |
| O (9) | 2e | 1 | 0.1857 | 0.9242 | 0.8883 |
| O (10) | 2e | 1 | 0.0574 | 0.7170 | 0.7575 |
| O (11) | 2e | 1 | 0.2373 | 0.4401 | 0.8499 |
| O (12) | 2e | 1 | 0.9824 | 0.5410 | 0.8903 |
| O (13) | 2e | 1 | 0.5077 | 0.1199 | 0.0748 |
| O (14) | 2e | 1 | 0.3810 | 0.3468 | 0.1790 |
| O (15) | 2e | 1 | 0.2394 | 0.9753 | 0.0925 |
| O (16) | 2e | 1 | 0.4782 | 0.8827 | 0.2029 |
| S (1) | 2e | 1 | 0.6137 | 0.7730 | 0.3810 |
| S (2) | 2e | 1 | 0.8494 | 0.3005 | 0.6157 |
| S (3) | 2e | 1 | 0.1158 | 0.6531 | 0.8467 |
| S (4) | 2e | 1 | 0.4019 | 0.0871 | 0.1366 |

Unit cell: monoclinic $P2(3)$; $a$ = 8.55092 Å, $b$ = 4.83219 Å, $c$ = 15.97060 Å, $\beta$ = 99.4°